\documentclass[aps,prl,superscriptaddress,twocolumn,a4paper]{revtex4}

\usepackage{graphicx,graphics,epsfig}   
\usepackage{dcolumn}    
\usepackage{bm}         
\usepackage{amsmath}    
\usepackage{verbatim}   
\usepackage{color}      
\usepackage{subfigure}  
\usepackage{times,natbib}
\usepackage{amsmath,amsfonts,amssymb,graphics,graphicx,epsfig,color,times,natbib}

\newcommand{\be}{\begin{equation}}
\newcommand{\ee}{\end{equation}}
\newcommand{\ba}{\begin{eqnarray}}
\newcommand{\ea}{\end{eqnarray}}
\newcommand{\ban}{\begin{eqnarray*}}
\newcommand{\ean}{\end{eqnarray*}}

\newcommand{\ket}[1]{\mbox{$ | #1 \rangle $}}

\definecolor{nblue}{rgb}{0.2,0.2,0.7}
\definecolor{ngreen}{rgb}{0.2,0.6,0.2}
\definecolor{nred}{rgb}{0.8,0.2,0.2}
\definecolor{nblack}{rgb}{0,0,0}

\renewcommand{\S}{\mathcal{S}}
\newcommand{\unit}{\mathbf{1}}

\newcommand{\proj}[1]{\mbox{$ | #1 \rangle\!\langle #1 | $}}
\newcommand{\mea}{\{\vec{a}_x,\vec{b}_y\}}

\begin{document}

\title{Guaranteed violation of a Bell inequality without aligned reference frames or
calibrated devices}

\author{Peter Shadbolt}
\affiliation{Centre for Quantum Photonics, H. H. Wills Physics Laboratory \&
Department of Electrical and Electronic Engineering, University of Bristol, Merchant
Venturers Building, Woodland Road, Bristol, BS8 1UB, UK}
\author{Tam\'as V\'ertesi}
\affiliation{Institute of Nuclear Research of the Hungarian Academy of Sciences,
H-4001 Debrecen, P.O. Box 51, Hungary.}
\author{Yeong-Cherng Liang}
\affiliation{Group of Applied Physics, University of Geneva, CH-1211 Geneva 4,
Switzerland}
\author{Cyril Branciard}
\affiliation{School of Mathematics and Physics, The University of Queensland, St
Lucia, QLD 4072, Australia}
\author{Nicolas Brunner}
\email{n.brunner@bristol.ac.uk}
\affiliation{H.H. Wills Physics Laboratory, University of Bristol, Tyndall Avenue,
Bristol, BS8 1TL, United Kingdom}
\author{Jeremy L. O'Brien}
\affiliation{Centre for Quantum Photonics, H. H. Wills Physics Laboratory \&
Department of Electrical and Electronic Engineering, University of Bristol, Merchant
Venturers Building, Woodland Road, Bristol, BS8 1UB, UK}

\date{\today}

\begin{abstract}
Bell tests---the experimental demonstration of a Bell inequality violation---are
central to understanding the foundations of quantum mechanics, underpin quantum
technologies, and are a powerful diagnostic tool for technological developments in
these areas. To date, Bell tests have relied on careful calibration of the measurement devices and alignment of a
shared reference frame between the two parties---both technically demanding tasks in general. 
Surprisingly, we show that neither of these operations are necessary, violating
Bell inequalities with near certainty with (i) unaligned, but calibrated, measurement devices, and (ii) uncalibrated and unaligned devices.
We demonstrate generic quantum nonlocality with randomly chosen local measurements on a singlet state of two photons 
implemented with reconfigurable integrated optical waveguide circuits based on voltage-controlled
phase shifters.
The observed results demonstrate the robustness of our schemes to imperfections and
statistical noise. 
This new approach is likely to have important applications in both fundamental science
and in quantum technologies, including device independent quantum key distribution. 
\end{abstract}

\maketitle

Nonlocality is arguably among the most striking aspects of quantum mechanics,
defying our intuition about space and time in a dramatic way \cite{bell}. Although
this feature was initially regarded as evidence of the incompleteness of the
theory~\cite{EPR}, there is today overwhelming experimental evidence that nature is
indeed nonlocal \cite{BellExp}. Moreover, nonlocality plays a central role in
quantum information science, where it proves to be a powerful resource, allowing,
for instance, for the reduction of communication complexity \cite{CC} and for
device-independent information processing~
\cite{DI,randomness1,randomness2,rabelo}.

\begin{figure}[t!]
   \includegraphics[width=8.5cm]{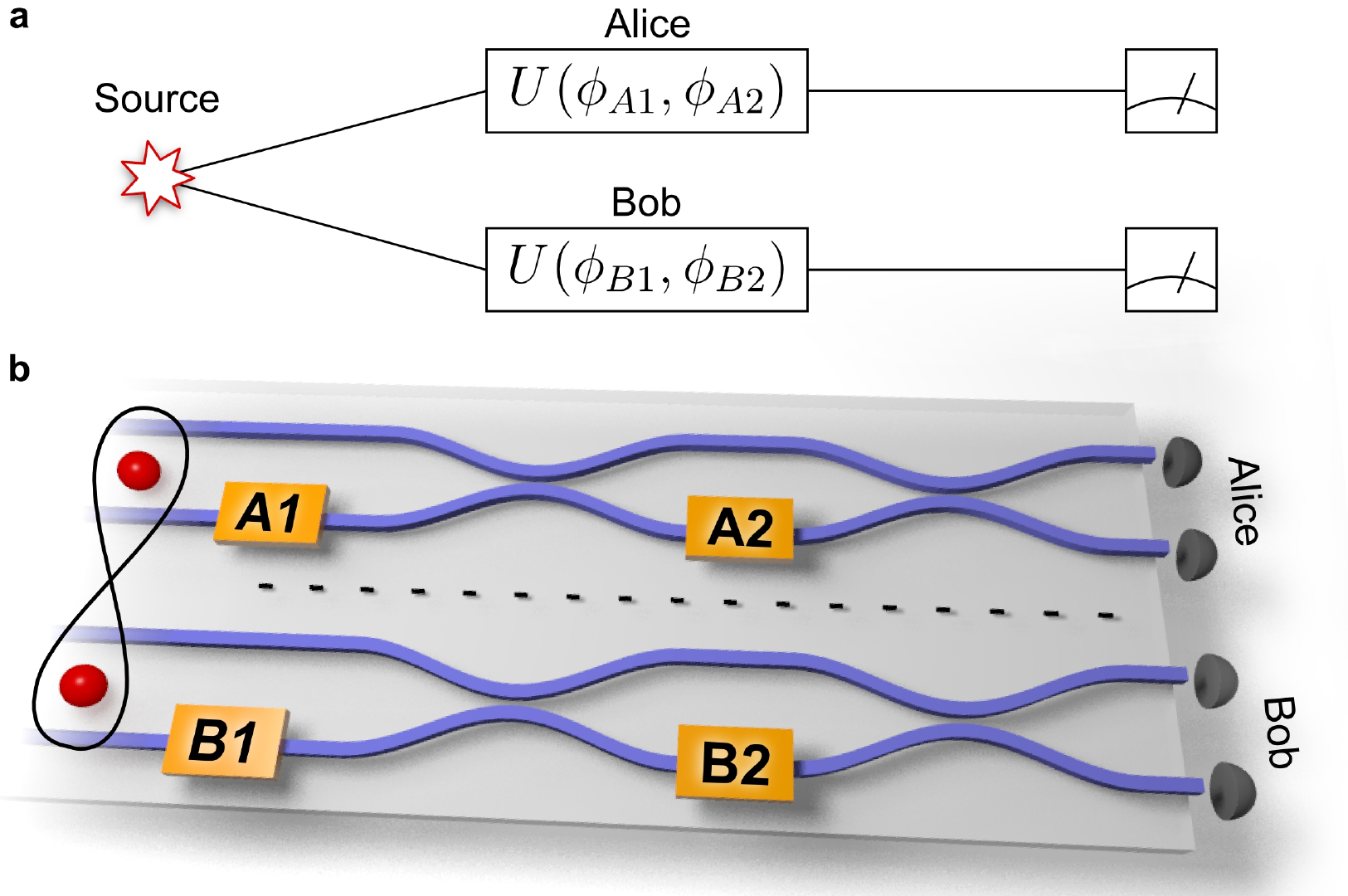}
\vspace{-0.1in}
   \caption{\label{fig:chip}  {\bf Bell violations with random measurements.} (a)
Schematic representation of a Bell test:. (b) Schematic of the integrated
waveguide chip used to implement the new schemes described here. Alice and Bob's
measurement circuits consist of waveguides to encode photonic qubits, directional
couplers that implement Hadamard-like operations, thermal phase shifters to
implement arbitrary measurements and detectors.} 
\end{figure}

In a quantum Bell test, two (or more) parties perform local measurements on an entangled quantum state,
Fig.~\ref{fig:chip}(a). After accumulating enough data, both parties can compute
their joint statistics and assess the presence of nonlocality by checking for the
violation of a Bell inequality. Although entanglement is necessary for obtaining
nonlocality it is not sufficient. First, there exist some mixed entangled states
that can provably not violate any Bell inequality since they admit a local
model~\cite{werner}. Second, even for sufficiently entangled states, one needs
judiciously chosen measurement settings~\cite{Y.C.Liang:PRA:042103}. Thus although
nonlocality reveals the presence of entanglement in a device-independent way, that
is, irrespectively of the detailed functioning of the measurement devices, one
generally considers carefully calibrated and aligned measuring devices
in order to obtain a Bell inequality violation. This in general amounts to having
the distant parties share a common reference frame and well calibrated devices.
Although this assumption is typically made implicitly in theoretical works, establishing a common reference frame, as well
as aligning and calibrating measurement devices in experimental situations are never
trivial issues. It is therefore an interesting and important question whether such
requirements can be dispensed with.

It was recently shown~\cite{RandBIV:2010} that, for Bell tests performed in the
absence of a shared reference frame, i.e., using randomly chosen measurement
settings, the probability of obtaining quantum nonlocality can be significant. For
instance, considering the simple Clauser-Horne-Shimony-Holt (CHSH)
scenario~\cite{CHSH}, randomly chosen measurements on the singlet state lead to a
violation of the CHSH inequality with probability of $\sim 28\%$; moreover this
probability can be increased to $\sim 42\%$ by considering unbiased measurement
bases. The generalization of these results to the multipartite case were considered in
Refs.~\cite{RandBIV:2010,RandBIV:2011}, as well as as schemes based on decoherence-free subspaces \cite{cabello}.
Although these works demonstrate that nonlocality can be a relatively common feature
of entangled quantum states and random measurements, it is of
fundamental interest and practical importance to establish whether Bell inequality violation can be ubiquitous.

Here we demonstrate that nonlocality is in fact a far more generic feature than
previously thought, violating CHSH inequalities without a shared frame of reference, and even with
uncalibrated devices, with near-certainty. We first show that whenever two parties perform
three mutually unbiased (but randomly chosen) measurements on a maximally entangled
qubit pair, they obtain a Bell inequality violation with certainty---a scheme that
requires no common reference frame between the parties, but only a local calibration
of each measuring device. We further show that when all measurements are chosen at
random (\emph{i.e.}, calibration of the  devices is not necessary anymore), although
Bell violation is not obtained with certainty, the probability of obtaining
nonlocality rapidly increases towards one as the number of different local
measurements increases. We perform these random measurements on 
the singlet state of two photons using a reconfigurable integrated waveguide
circuit, based on voltage-controlled phase shifters. The data confirm the near-unit
probability of violating an inequality as well as the robustness of the scheme to
experimental imperfections---in particular the non-unit visibility of the entangled
state---and statistical uncertainty. 
These new schemes exhibit a surprising robustness of the observation of nonlocality
that is likely to find important applications in diagnostics of quantum devices
(\emph{e.g..} removing the need to calibrate the reconfigurable circuits used here)
and quantum information protocols, including device independent quantum key
distribution \cite{DI} and other protocols based on quantum nonlocality \cite{randomness1,randomness2,rabelo} and quantum steering \cite{onesided}.

\vspace{8pt}
\noindent{\bf Results}

\noindent{\bf \emph{Bell test using random measurement triads.}} Two distant
parties, Alice and Bob, share a Bell state. Here we will focus on the singlet state
\begin{equation}\label{Eq:Singlet}
        \ket{\Psi^-}=\frac{1}{\sqrt{2}}\left(\ket{0}_A\ket{1}_B-\ket{1}_A\ket{0}_B\right),
\end{equation}
though all our results can be adapted to hold for any two-qubit maximally entangled
state. Let us consider a Bell scenario in which each party can perform 3 possible
qubit measurements labeled by the Bloch vectors $\vec{a}_x$ and $\vec{b}_y$
($x,y=1,2,3)$, and where each measurement gives outcomes $\pm1$. After sufficiently
many runs of the experiment, the average value of the product of the measurement
outcomes, i.e. the correlators $E_{xy}=-\vec{a}_x\cdot\vec{b}_y$, can be estimated
from the experimental data. In this scenario, it is known that all local measurement
statistics must satisfy the CHSH inequalities:
\begin{equation}\label{chsh}
        \left| E_{xy}+E_{xy'}+E_{x'y}-E_{x'y'}\right|\le 2,
\end{equation}
and their equivalent forms where the negative sign is permuted to the other terms
and for different pairs $x,x'$ and $y,y'$; there are, in total, 36 such
inequalities. 

Interestingly, it turns out that whenever the measurement settings are
unbiased, i.e. $\vec{a}_{x}\cdot\vec{a}_{x'}=\delta_{x,x'}$ and
$\vec{b}_{y}\cdot\vec{b}_{y'}=\delta_{y,y'}$, then at least one of the above CHSH
inequalities {\it must be violated} --- except for the case where the orthogonal triads (from now on simply referred to as triads) are
perfectly aligned, i.e. for each $x$, there is a $y$ such that
$\vec{a}_{x}=\pm\vec{b}_{y}$. 
Therefore, a generic random choice of unbiased measurement settings --- where the
probability that Alice and Bob's settings are perfectly aligned is zero (for
instance if they share no common reference frame) --- will \emph{always} lead to the
violation of a CHSH inequality. 

\emph{Proof.}
Assume that $\{\vec a_x\}$ and $\{\vec b_y\}$ are orthonormal bases. Since the
correlators of the singlet state have the simple scalar product form $E_{xy} =
-\vec{a}_x\cdot\vec{b}_y$, the matrix
\ba\label{matrix}
{\cal E} = \left(
\begin{array}{ccc}
E_{11} & E_{12} & E_{13} \\
E_{21} & E_{22} & E_{23} \\
E_{31} & E_{32} & E_{33}
\end{array}
\right)
\ea
contains (in each column) the coordinates of the three vectors $-\vec b_y$, written
in the basis $\{\vec a_x\}$.

By possibly permuting rows and/or columns, and by possibly changing their signs (which
corresponds to relabeling Alice and Bob's settings and outcomes), we can assume,
without loss of generality, that $E_{11}, E_{22} > 0$ and that $E_{33} > 0$ is the
largest element (in absolute value) in the matrix ${\cal E}$. Noting that $\vec{b}_3
= \pm \vec{b}_1 \times \vec{b}_2$ and therefore $|E_{33}| = |E_{11} E_{22} - E_{12}
E_{21}|$, these assumptions actually imply $E_{33} = E_{11} E_{22} - E_{12} E_{21}
\geq E_{11} , E_{22} , |E_{12}|, |E_{21}|$ and $E_{12} E_{21} \leq 0$; we will
assume that $E_{12} \leq 0$ and $E_{21} \geq 0$ (one can multiply both the $x=2$ row
and the $y=2$ column by -1 if this is not the case).

With these assumptions, $(E_{11} + E_{21}) \max[-E_{12},E_{22}] \geq E_{11} E_{22} -
E_{12} E_{21} = E_{33} \geq \max[-E_{12},E_{22}]$, and by dividing by
$\max[-E_{12},E_{22}] > 0$, we get $E_{11} + E_{21} \geq 1$. One can show in a
similar way that $-E_{12} + E_{22} \geq 1$. Adding these last two inequalities, we
obtain
\ba
E_{11} + E_{21} - E_{12} + E_{22} \geq 2 \,. \label{ineq_proof}
\ea
Since ${\cal E}$ is an orthogonal matrix, one can check that equality is obtained
above (which requires that both $E_{11} + E_{21} = 1$ and $-E_{12} + E_{22} = 1$) if
and only if $\vec{a}_1 = \vec{b}_1$, $\vec{a}_2 = \vec{b}_2$ and $\vec{a}_3 =
\vec{b}_3$. 
Therefore, if the two sets of mutually unbiased measurement settings $\{\vec a_x\}$
and $\{\vec b_y\}$ are not aligned, then inequality (\ref{ineq_proof}) is strict: a
CHSH inequality is violated. Numerical evidence suggests that the above construction
always gives the largest CHSH
violation obtainable from the correlations \eqref{matrix}. $\hfill\blacksquare$

\begin{figure}[h!]
    \includegraphics[height=5.5cm,width=9cm]{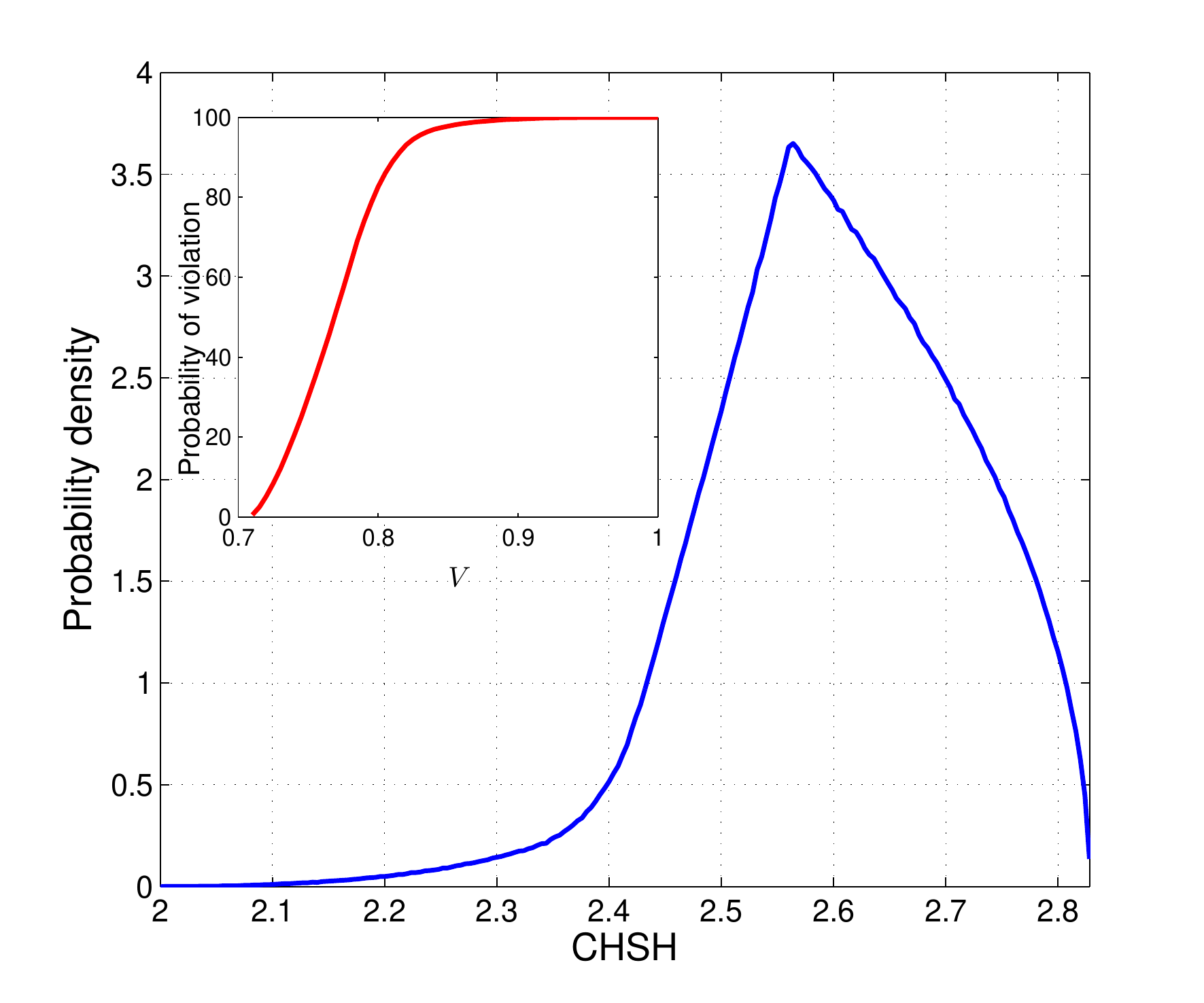}
    \caption{\label{Fig:OrthTriads3322C-3D}  {\bf Bell tests using random
measurement triads (theory).} Distribution of the (maximum) CHSH violations for
uniformly random measurement triads on a singlet state. The inset shows the
probability of obtaining a CHSH violation as a function of the visibility $V$ of
the Werner state; this probability is obtained by integrating the distribution
of CHSH violations (main graph) over the interval $[2/V,2\sqrt{2}]$.
}
\end{figure}

While the above result shows that a random choice of measurement triads will lead to
nonlocality with certainty, we still need to know how these CHSH violations are
distributed; that is, whether the typical violations will be rather small or large.
This is crucial especially for experimental implementations, since in practice,
various sources of imperfections will reduce the strength of the observed
correlations. Here we consider two main sources of imperfections: limited
visibility of the entangled state, and finite statistics. 

First, the preparation of a pure singlet state is impossible experimentally and it
is thus desirable to understand the effect of depolarizing noise, giving rise to a
reduced visibility $V$: 
\begin{equation}\label{werner}
        \ket{\Psi^-}\rightarrow \rho_V = V \proj{\Psi^-}+(1-V)\frac{\unit}{4}.
\end{equation}
This, in turn, results in the decrease of the strength of correlations by a factor
$V$. In particular, when $V\leq1/\sqrt{2}$, the state~\eqref{werner} ceases to
violate the CHSH inequality. States $\rho_V$ are known as Werner states
\cite{werner}.

Second, in any experiment the correlations are estimated from a finite set of data,
resulting in an experimental uncertainty. To take into account this finite-size
effect, we will consider a shifted classical bound $\S \geq 2$ of the CHSH
expression \eqref{chsh} such that an observed correlation is only considered to give
a conclusive demonstration of nonlocality if $\text{CHSH}>\S$. Thus, if the CHSH
value is estimated experimentally up to a precision of $\delta$, then considering a
shifted classical bound of $\S = 2+\delta$ ensures that only statistically
significant Bell violations are considered.

We have estimated numerically the distribution of the CHSH violations (the maximum
of the left-hand-side of~\eqref{chsh} over all $x,x',y,y'$) for uniformly random
measurement triads on the singlet state (see
Fig.~\ref{Fig:OrthTriads3322C-3D}). Interestingly, typical violations are quite
large; the average CHSH value is $\sim 2.6$, while only $\sim0.3\%$ of the
violations are below $2.2$. Thus this phenomenon of generic nonlocality is very
robust against the effect of finite statistics and of limited visibility,
even in the case where both are combined. For instance, even after raising the
cutoff to $\S= 2.1$ and decreasing the singlet visibility to $V=0.9$, our numerical
simulation shows that the probability of violation is still greater than 98.2\% (see
Fig~\ref{Fig:OrthTriads3322C-3D}).

\vspace{8pt}
\noindent{\bf \emph{Bell tests using completely random measurements.}} Although
performing unbiased measurements does not require the spatially separated parties to
share a common reference frame, it still requires each party to have good control of
the local measurement device. Clearly, local alignment errors (that is, if the
measurements are not exactly unbiased) will reduce the probability of obtaining
nonlocality. In practice the difficulty of correctly aligning the local measurement
settings depends on the type of encoding that is used. For instance, using the
polarization of photons, it is rather simple to generate locally a measurement
triad, using wave-plates. However, for other types of encoding, generating unbiased
measurements might be much more complicated (see experimental part below).

This leads us to investigate next the case where all measurement directions $\mea$
are chosen randomly and independently. For simplicity, we will focus here on the case where all
measurements are chosen according to a uniform distribution on the Bloch sphere.
Although this represents a particular choice of distribution, we believe that most
random distributions that will naturally arise in an experiment will lead to
qualitatively similar results, as indicated by our experimental results. 

\begin{figure}
    \includegraphics[height=5.5cm,
width=9cm]{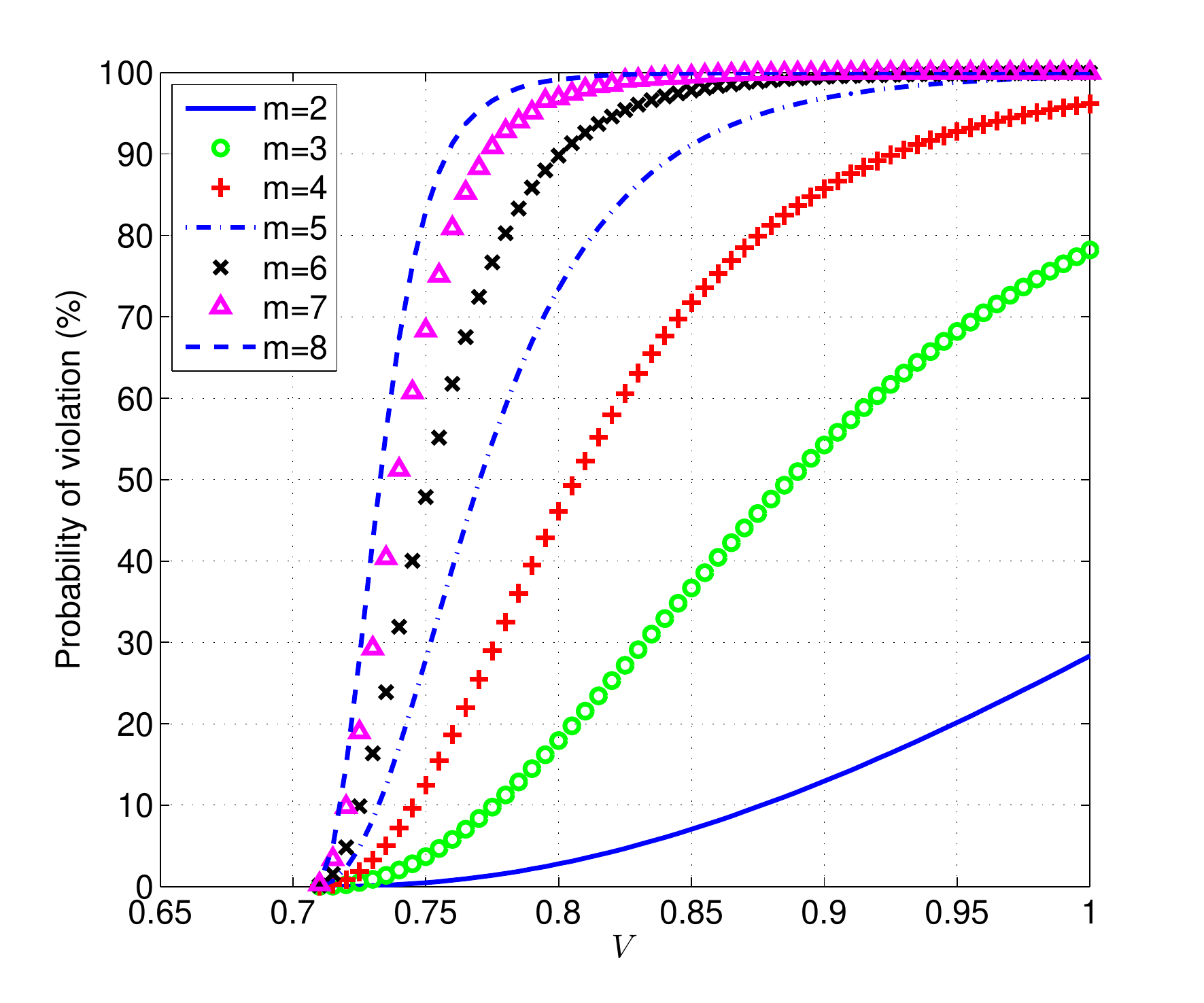}
    \caption{\label{Fig:ROM}  {\bf Bell tests using completely random measurements
(theory).} Plot of the probability of Bell violation as a function of the
visibility $V$ of the Werner state, for different numbers $m$ of (completely
random) measurements per party.}
\end{figure}

\begin{figure*}[t!]
\centering
\subfigure{\raisebox{0cm}{\includegraphics[width=85mm]{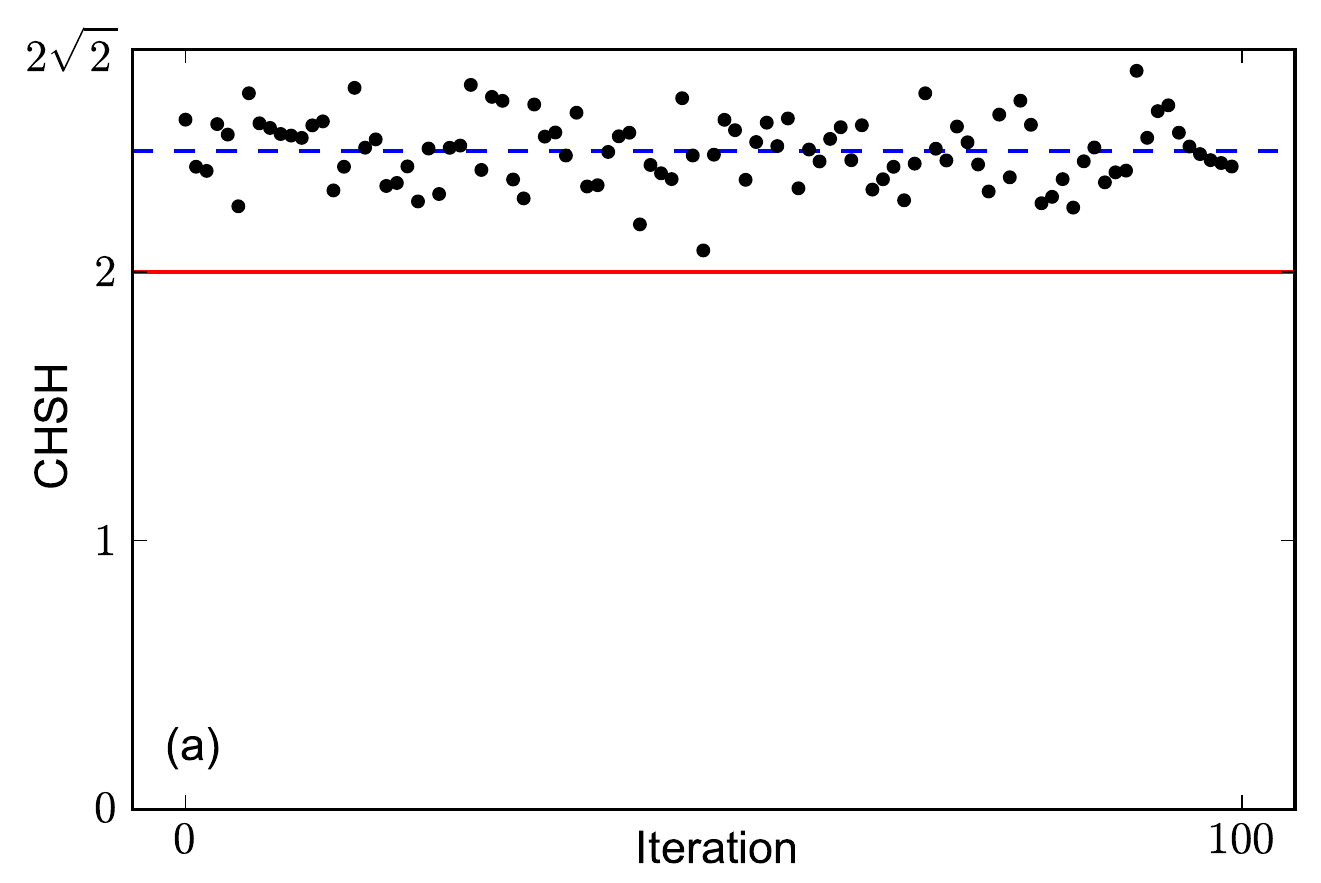}}}
\subfigure{\raisebox{-3mm}{\includegraphics[width=85mm]{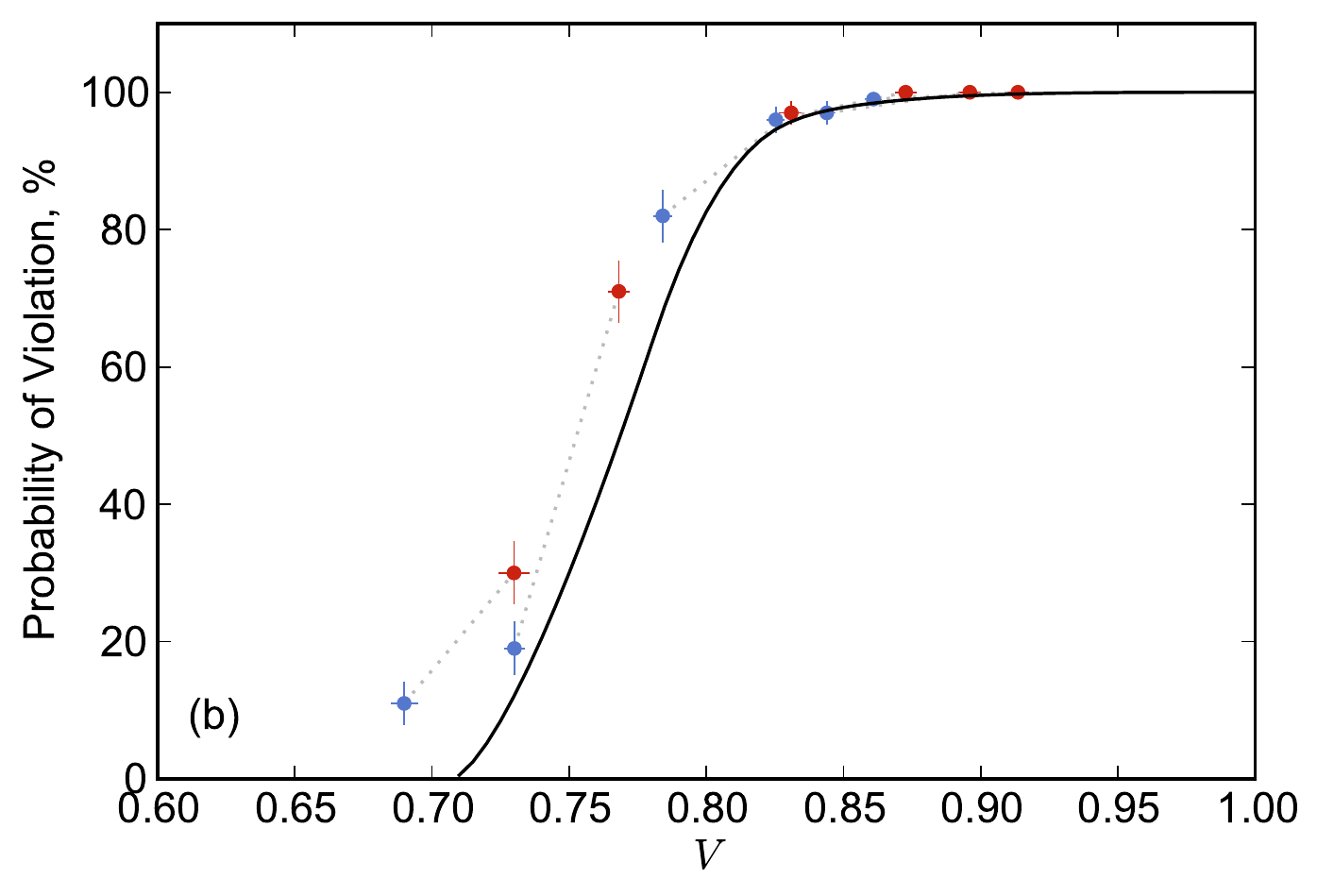}}}\vspace{-2mm}
\caption{ {\bf Bell tests requiring no shared reference frame.} Here we
perform Bell tests on a two-qubit Bell state, using randomly chosen measurement
triads. Thus our experiment requires effectively no common reference frame between
Alice and Bob. (a) 100 successive Bell tests; in each iteration, both Alice and Bob
use a randomly-chosen measurement triad. For each iteration, the maximal CHSH value
is plotted (black points). In all iterations, we get a CHSH violation; the red line
indicates the local bound (CHSH=2). The smallest CHSH value is $\sim 2.1$, while the
mean CHSH value (dashed line) is $\sim 2.45$. This leads to an estimate of the
visibility of $V=\frac{2.45}{2.6}\simeq 0.942$, to be compared with $0.913\pm0.004$
obtained by maximum likelihood quantum state tomography \cite{maxlike}. This slight
discrepancy is due to the fact that our entangled state is not exactly of the form
of a Werner state. Error bars, which are too small to draw, were estimated using a
Monte Carlo technique, assuming Poissonian photon statistics. (b) The experiment of
(a) is repeated for Bell states with reduced visibility, illustrating the robustness
of the scheme. Each point shows the probability of CHSH violation estimated
using 100 trials. Uncertainty in probability is estimated as the standard error.
Visibility for each point is estimated by maximum-likelihood quantum state
tomography, where the error bar is calculated using a Monte Carlo approach, again
assuming Poissonian statistics. Red points show data corrected for accidental
coincidences (see Methods), the corresponding uncorrected data is shown in blue.
Again, since our state is not exactly of the Werner form, we get slightly higher
probabilities of CHSH violation than expected. \label{fig:triads_expt}}
\end{figure*}

We thus now consider a Bell test in which Alice and Bob share a singlet, and each
party can use $m$ possible measurement settings, all chosen randomly and uniformly
on the Bloch sphere. We estimated numerically the probability of getting a Bell
violation as a function of the visibility $V$ [of the state~\eqref{werner}] for
$m=2,\ldots,8$; see Fig.~\ref{Fig:ROM}. Note that for $m\geq4$, additional Bell
inequalities appear \cite{gisin}; we have checked however, that ignoring
these inequalities and considering only CHSH leads to the same results up to a very
good approximation. Fig. \ref{Fig:ROM} clearly shows that the chance of
finding a nonlocal correlation rapidly increases with the number of settings $m$.
Intuitively, this is because when choosing an increasing number of measurements at
random, the probability that at least one set of four measurements (2 for Alice and
2 for Bob) violates the CHSH inequality increases rapidly. For example, with $m=3$
settings, this probability is $78.2\%$  but with $m=4$, it is already 96.2\%, and
for $m=5$ it becomes 99.5\%. Also, as with the case of unbiased measurements, the
probability of violation turns out to be highly robust against depolarizing noise;
for instance, for $V=0.9$ and $m\geq 5$, there is still at least 96.9\% chance of finding a subset $\{\vec{a}_x,\vec{a}_{x'},\vec{b}_y,\vec{b}_{y'}\}$ among our randomly chosen measurements that gives nonlocal correlations.

\vspace{8pt}
\noindent{\bf \emph{Measurement devices.}} 
We use the device shown in Fig.~\ref{fig:chip}(b) to implement Alice and Bob's
random measurements on an entangled state of two photons. The path encoded singlet
state is generated from two unentangled photons using an integrated waveguide
implementation \cite{politi1,politi2} of a nondeterministic
\textsc{cnot} gate \cite{cnot1,cnot2,cnot3} to enable
deliberate introduction of mixture. This state is then shared between Alice and Bob
who each have a Mach-Zehnder (MZ) interferometer, consisting of two directional couplers
(equivalent to beamsplitters) and variable phase shifters ($A_{1,2}$ and $B_{1,2}$)
and single photon detectors. This enables Alice and Bob to independently make a projective
measurement in any basis by setting their phase shifters to the required values \cite{matthews,pete}.
The effect of the first phase-shifter is a rotation around the $Z$ axis of the Bloch
sphere ($R_{Z}(\phi_1)=e^{-i\phi_1 \sigma_{Z}/2}$); since each directional coupler implements a
Hadamard-like operation ($H'=e^{i\pi/2} e^{-i\pi\sigma_{Z}/4} H
e^{-i\pi\sigma_{Z}/4}$, where $H$ is the usual Hadamard gate), the effect of the
second phase shifter is a rotation around the $Y$ axis ($R_{Y}(\phi_2)=e^{-i\phi_2
\sigma_{Y}/2}$). Overall the MZ interferometer and phase shifters implement the
unitary transformation $U(\phi_1,\phi_2) = R_{Y}(\phi_2) R_{Z}(\phi_1)$, which
enables projective measurement in any qubit basis when combined with a final
measurement in the logical ($Z$) basis using avalanche photodiode single photon
detectors (APDs).

Each thermal phase shifter is implemented as a resistive element, lithographically
patterned onto the surface of the waveguide cladding. Applying a voltage $v$ to the
heater has the effect of locally heating the waveguide, thereby inducing a small
change in refractive index $n$ ($dn / dT \approx 1 \times 10^{-5}K$) which manifests
as a phase shift in the MZ interferometer. There is a nonlinear relationship between
the voltage applied and the resulting phase shift $\phi(v)$, which is generally well
approximated by a quadratic relation of the form
\begin{equation}
\phi(v) = \alpha + \beta v^2 .
\label{phiv}
\end{equation}

In general, each heater must be characterized individually, \emph{i.e.} by
estimating the function $\phi(v)$. This is achieved by measuring single-photon
interference fringes from each heater. The parameter $\alpha$ can take any value
between 0 and $2\pi$ depending on the fabrication of the heater, while typically
$\beta\sim0.15 \frac{\text{rad}}{V^2}$ \cite{matthews,pete}. 
For any desired phase, the corrected voltage can then be determined. This operation
is necessary both for state tomography, and for implementing random measurement
triads. 
In contrast, this calibration can be dispensed with when implementing completely
random measurements.
Indeed this represents the central point of this experiment, which requires no a
priori calibration of the devices. Thus, 
In this case we simply choose random voltages from a uniform distribution, in the
range $[0\text{V} , 7\text{V}]$, which is adequate to address phases in the range $0
\le \phi \le 2\pi$---\emph{i.e.} no a priori calibration of Alice and Bob's devices
is necessary.

\vspace{8pt}\noindent{\bf \emph{Experimental violations with random measurement triads.}}
We first investigate the situation in which Alice and Bob both use 3 orthogonal
measurements. We generate randomly chosen measurement triads using a pseudo-random
number generator. Having calibrated the phase/voltage relationship of the
phase shifters, we then apply the corresponding voltages on the chip. For each pair of measurement
settings, the coincidence counts between all of the 4 pairs of APDs are then
measured for a fixed amount of time---the typical number of simultaneous photon
detection coincidences is $\sim1$ kHz. From these data we compute the maximal CHSH
value as detailed above. This entire procedure is then repeated 100 times. The
results are presented in Fig.~\ref{fig:triads_expt}a, where accidental coincidences,
arising primarily from photons originating from different down-conversion events,
which are measured throughout the experiment, have been subtracted from the data
(the raw data, as well as more details on accidental events, can be found in the
Methods). Remarkably, all 100 trials lead to a clear CHSH violation; the
average CHSH value we observe is $\sim 2.45$, while the smallest measured value is
$\sim 2.10$.

We next investigate the effect of decreasing the visibility of the singlet state: By
deliberately introducing a temporal delay between the two photons arriving at the
\textsc{cnot} gate, we can increase the degree of distinguishability between the two
photons. Since the photonic \textsc{cnot} circuit relies on quantum interference
\cite{HOM} a finite degree of distinguishability between the photons
results in this circuit implementing an incoherent mixture of the CNOT operation and
the identity operation \cite{ob_prl}. 
By gradually increasing the delay we can create states $\rho_V$ with decreasing
visibilities. For each case, the protocol described above is repeated, which allows
us to estimate the average CHSH value (over 100 trials). For each case we also
estimate the visibility via maximum likelihood quantum state tomography. Figure
\ref{fig:triads_expt}b clearly demonstrates the robustness of our scheme, in good
agreement with theoretical predictions: a considerable amount of mixture must be
introduced in order to get an average CHSH value below 2.

Together these results show that large Bell violations can be obtained without a
shared reference frame even in the presence of considerable mixture.

\begin{figure}[t!]
   \includegraphics[width=8.7cm]{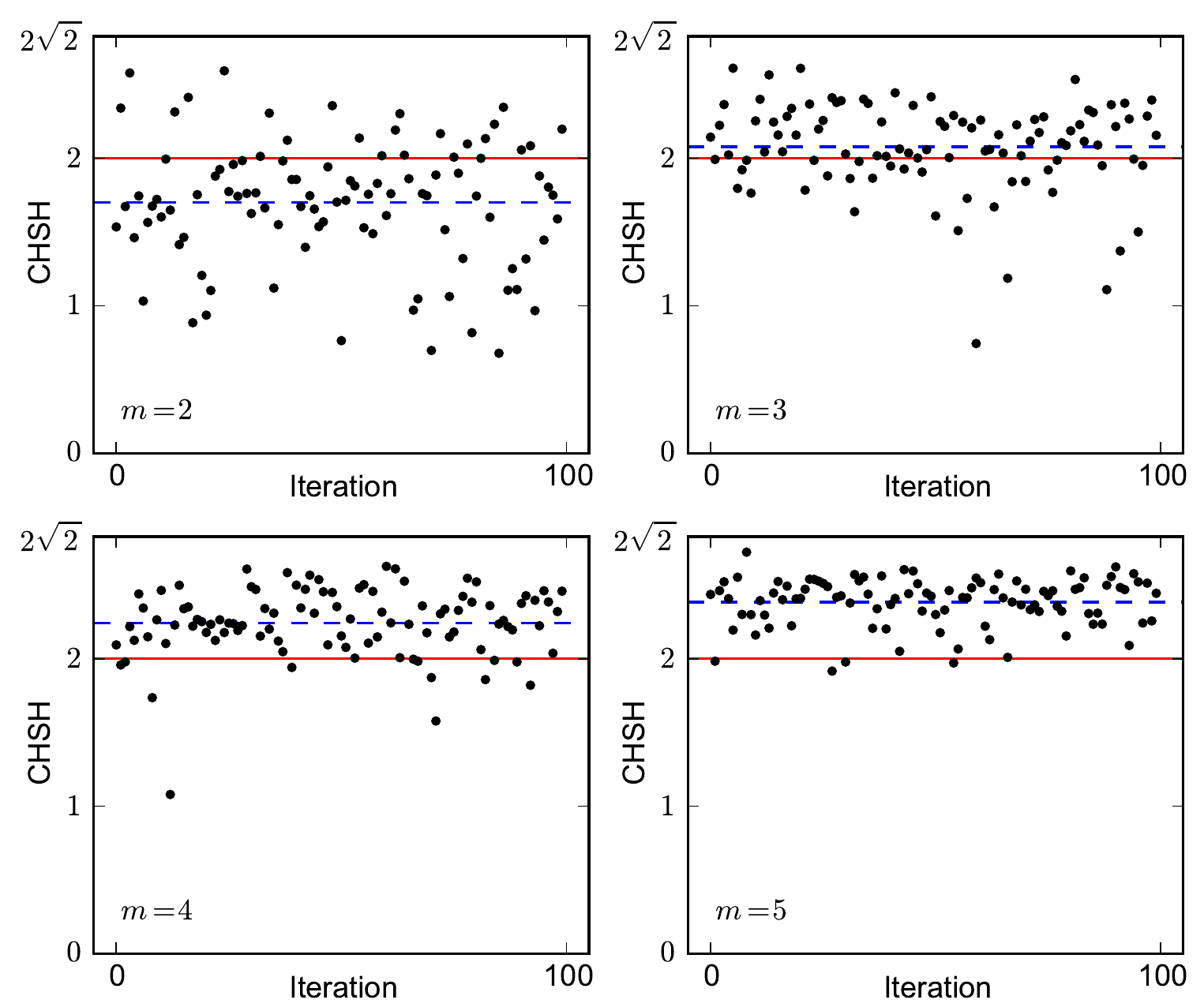}
    \caption{\label{fig:random_voltages_corrected}  {\bf Experimental Bell tests
using uncalibrated devices.} We perform Bell tests on a two-qubit Bell
state, using uncalibrated measurement interferometers, that is, using
randomly-chosen voltages. For $m=2,3,4,5$ local measurement settings, we perform
100 trials (for each value of $m$). As the number of measurement settings $m$ increases, the probability
of obtaining a Bell violation rapidly approaches one. For $m\geq 3$, the average
CHSH value (dashed line) is above the local bound of CHSH=2 (red line). Error
bars, which are too small to draw, were estimated by a Monte Carlo technique,
assuming Poissonian statistics. Data has been corrected for accidentals (see
Methods).}
\end{figure}

\vspace{8pt}\noindent{\bf \emph{Experimental violations with completely random measurements.}}
We now investigate the case where all measurements are chosen at random. The
procedure is similar to the first experiment, but we now apply voltages chosen
randomly from a uniform distribution, and independently for each measurement
setting. Thus our experiment requires 
no calibration of the measurement MZ interferometers (\emph{i.e.}~the
characterization of the phase-voltage relation), which is generally a cumbersome
task. By increasing the number of measurements performed by each party ($m=2,3,4,5$), we
obtain CHSH violations with a rapidly increasing probability, see
Fig.~\ref{fig:random_voltages_corrected}. For $m=5$, we find 95 out of 100 trials
lead to a CHSH violation. The visibility $V$ of the state used for this experiment
was measured using state tomography to be $0.869\pm0.003$, clearly demonstrating
that robust violation of Bell inequalities is possible for completely random
measurements. 

It is interesting to note that the relation between the phase and the applied
voltage is typically quadratic, see Eq.~(\ref{phiv}). 
Thus, by choosing voltages from a uniform distribution, the corresponding phase
distribution is clearly biased. Our experimental results indicate that this bias has
only a minor effect on the probability of obtaining nonlocality. 

\vspace{8pt}
\noindent{\bf{Discussion.}} 

\noindent Bell tests provide one of the most important paths to gaining insight into
the fundamental nature of quantum physics. The fact that they can be robustly
realised without the need for a shared reference frame or calibrated devices
promises to provide new fundamental insight. In the future it would be interesting
to investigate these ideas in the context of other Bell tests, for instance
considering other entangled states or in the multipartite situation (see \cite{joel}
for recent progress in this direction), as well as in the context of quantum reference frames \cite{review,costa}.

The ability to violate Bell inequalities with a completely uncalibrated device, as
was demonstrated here, has important application for the technological development
of quantum information science and technology: Bell violations provide an
unambiguous signature of quantum operation and the ability to perform such
diagnostics without the need to first perform cumbersome calibration of devices
should enable a significant saving in all physical platforms.  These ideas could be
particularly helpful for the characterization of entanglement sources without the
need for calibrated and aligned measurement devices.

Finally Bell violations underpin many quantum information protocols, and therefore,
the ability to realise them with dramatically simplified device requirements holds
considerable promise for simplifying the protocols themselves. For example, device
independent quantum key distribution \cite{DI} allows two parties to exchange a cryptographic key and, by checking for the violation of a Bell inequality, to guarantee its security without having a detailed knowledge of the devices used in the protocol. Such schemes,
however, do typically require precise control of the apparatus in order to obtain a
sufficiently large violation. In other words, although a Bell inequality violation
is an assessment of entanglement that is device-independent, one usually needs
carefully calibrated devices to obtain such a violation. The ability to violate Bell
inequalities without these requirements could dramatically simplify these
communication tasks. The implementation of protocols based on quantum steering \cite{onesided}
may also be simplified by removing calibration requirements.

\emph{Note added.} While completing this manuscript, we became aware of an
independent proof of our theoretical result on Bell tests with randomly chosen
measurement triads, obtained by Wallman and Bartlett \cite{joel}, after one of us
mentioned numerical evidence of this result to them.

\begin{acknowledgments}
We acknowledge useful discussions with Jonathan Allcock, Nicolas Gisin and Joel Wallman. We
acknowledge financial support from the UK EPSRC, ERC, QUANTIP, PHORBITECH, Nokia,
NSQI, the J\'anos Bolyai Grant of the Hungarian Academy of Sciences, the Swiss NCCR
"Quantum Photonics", and the European ERC-AG QORE. J.L.O'B. acknowledges a Royal
Society Wolfson Merit Award.
\end{acknowledgments}


\section{Methods}

\noindent\emph{Photon counting and accidentals.} In our experiments, we postselect
on successful operation of the linear-optical \textsc{cnot} gate by counting
coincidence events, that is, by measuring the rate of coincidental detection of
photon pairs. Single photons are first detected using silicon avalanche photodiodes
(APDs). Coincidences are then counted using a Field-Programmable Gate Array (FPGA)
with a time window of $\sim5$ns. We refer to these coincidence events as $\{ t_0^A,
t_0^B\}$.

\begin{figure}[b!]
   \includegraphics[width=8.2cm]{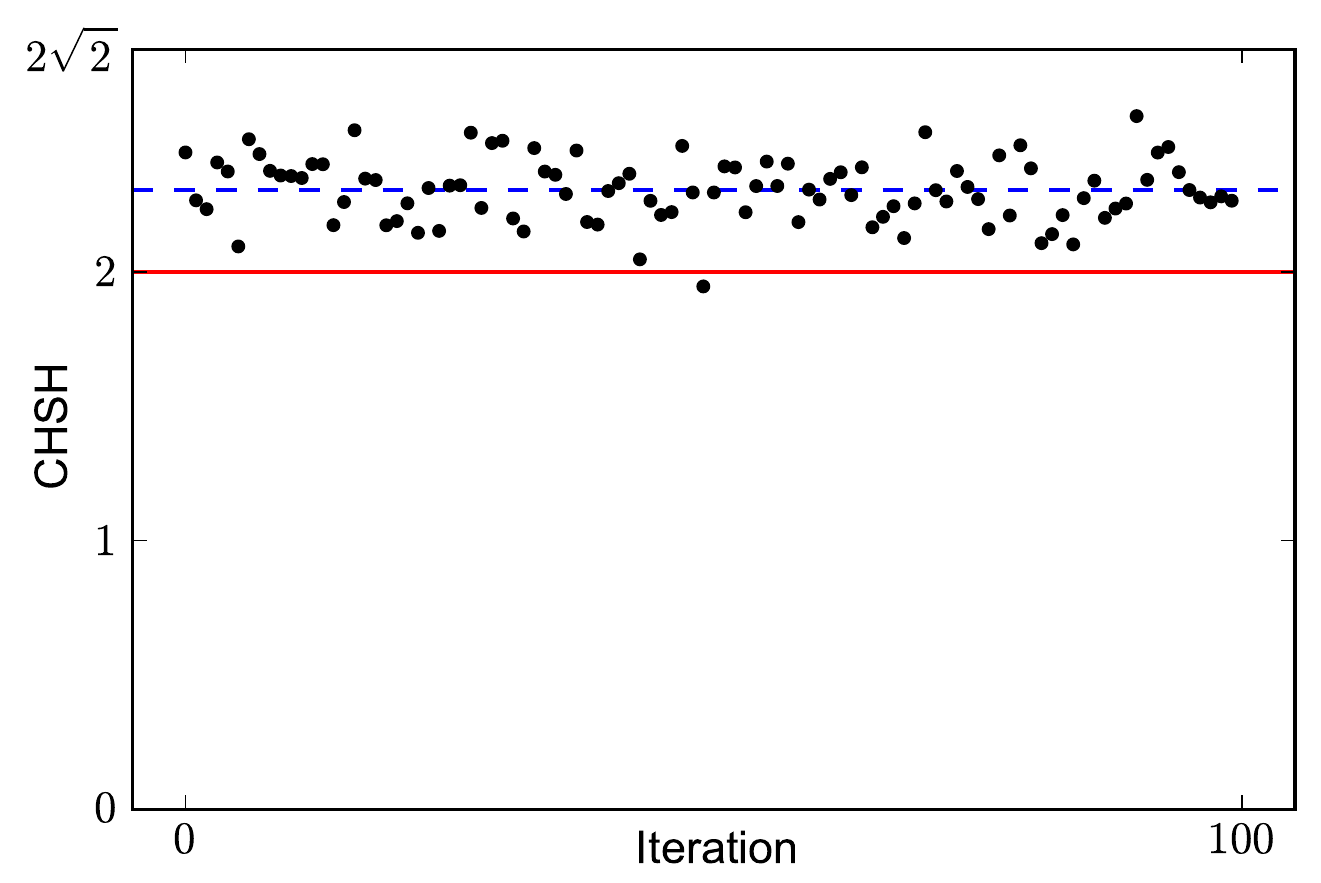}
    \caption{\label{fig:triads_expt_uncorrected}  {\bf Raw data of experimental Bell
tests requiring no shared reference frame.} This figure shows the raw data,
without correcting for accidental coincidences, of Fig.~4a. Here the average
CHSH value is 2.30 (dashed line), leading to an estimate of the visibility of
$V=\frac{2.3}{2.6}\simeq 0.885$, while the estimate from quantum state
tomography is $V=0.861\pm0.003$. Again, this discrepancy is due to the fact that
our entangled state is not exactly of the form of a Werner state. Error bars,
which are too small to draw, were estimated using a Monte-Carlo technique,
assuming Poissonian photon statistics.}
\end{figure}

\begin{figure}[b!]
   \includegraphics[width=8.5cm]{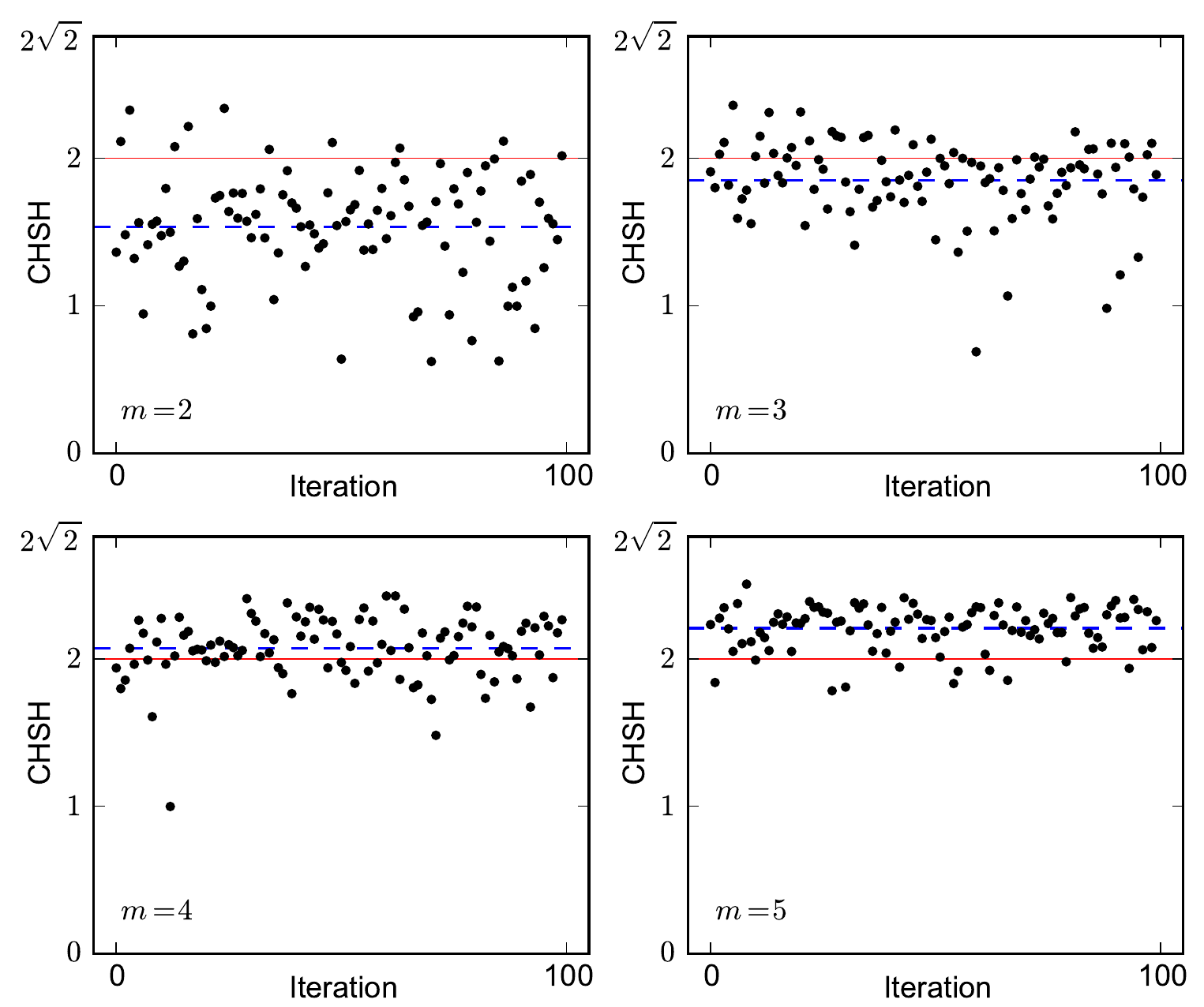}
    \caption{\label{fig:random_voltages_uncorrected}  {\bf Raw data of experimental
Bell tests using uncalibrated measurement interferometers (random
voltages).} This figure shows the raw data, without correcting for accidental
coincidences, of Fig.~5. Error bars are estimated by a Monte Carlo technique,
assuming Poissonian statistics. The visibility $V$ of the state used for this
experiment was measured using state tomography to be $0.804\pm0.003$.}
\end{figure} 
 
Accidental coincidences have two main contributions: first, from photons originating
from different down-conversion events arriving at the detectors within the time
window; second, due to dark counts in the detectors. Here we directly measure the
(dynamic) rate of accidental coincidences in real time, for the full duration of all
the experiments described here. To do so, for each pair of detectors we measure a
second coincidence count rate, namely $\{ t_0^A, t_1^B\}$, with $|t_1-t_0| =
30\text{ns}$. In order to do this, we first split (duplicate) the electrical TTL
pulse from each detector into two BNC cables. An electrical delay of 30ns is
introduced into one channel, and coincidences (i.e. at $\{ t_0^A, t_0^B\}$) are then
counted directly. Finally we obtain the corrected coincidence counts by subtracting
coincidence counts at $\{ t_0^A, t_1^B\}$ from the raw coincidence counts at $\{
t_0^A, t_0^B\}$. 

All experimental results presented in the main text have been corrected for
accidentals. Here we provide the raw data. Fig \ref{fig:triads_expt_uncorrected}
presents the raw data for Fig. \ref{fig:triads_expt}(a) while Fig.
\ref{fig:random_voltages_uncorrected} presents the raw data for Fig.
\ref{fig:random_voltages_corrected}. Notably, in
Fig.~\ref{fig:triads_expt_uncorrected}, corresponding to the case of randomly chosen
triads, all but one of the hundred trials feature a CHSH violation. The average
violation is now $\sim 2.3$.


\begin{thebibliography}{99}

\bibitem{bell} J. S. Bell, Physics (Long Island City, N.Y.) {\bf 1}, 195 (1964).

\bibitem{EPR} A. Einstein, B. Podolsky, and N. Rosen, Phys. Rev. {\bf 47}, 777 (1935).

\bibitem{BellExp} A.~Aspect, Nature {\bf 398}, 189 (1999).

\bibitem{CC} H. Buhrman, R. Cleve, S. Massar, and R. de~Wolf, Rev. Mod. Phys. {\bf 82}, 665 (2010).

\bibitem{DI} A. Ac\'in, N. Brunner, N. Gisin, S. Massar, S. Pironio, and V. Scarani,
Phys. Rev. Lett. {\bf 98}, 230501 (2007).

\bibitem{randomness1} S. Pironio \emph{et al}, Nature {\bf 464}, 1021 (2010). 

\bibitem{randomness2} R. Colbeck and A. Kent, J. Phys. A {\bf 44}, 095305 (2011).

\bibitem{rabelo} R. Rabelo, M. Ho, D. Cavalcanti, N. Brunner, and V. Scarani, Phys.
Rev. Lett. {\bf 107}, 050502 (2011).

\bibitem{werner} R.F. Werner, Phys. Rev. A {\bf 40}, 4277 (1989).

\bibitem{Y.C.Liang:PRA:042103} Y.-C.~Liang and A.~C.~Doherty,  \pra { \bf 75},
042103 (2007).

\bibitem{RandBIV:2010} Y.-C.~Liang, N.~Harrigan, S.~D.~Bartlett, and T.~G.~Rudolph,
    \prl { \bf 104}, 050401 (2010).

\bibitem{CHSH} J.~F.~Clauser, M.~A.~Horne, A.~Shimony and
        R.~Holt, \prl { \bf 23}, 880 (1969).

\bibitem{RandBIV:2011} J.~J.~Wallman, Y.-C.~Liang, and S.~D.~Bartlett, \pra { \bf
83}, 022110 (2011).

\bibitem{cabello} A. Cabello, Phys. Rev. Lett. {\bf 91}, 230403 (2003).

\bibitem{onesided} C. Branciard, E.G. Cavalcanti, S.P. Walborn,
V. Scarani, H.M. Wiseman, arXiv:1109.1435.

\bibitem{gisin} N. Gisin, arXiv:quant-ph/0702021.

\bibitem{politi1} A. Politi, M.J. Cryan, J.G. Rarity, S. Yu, and J.L. O'Brien, Science {\bf 320}, 646 (2008).

\bibitem{politi2} A. Politi, J.C.F. Matthews, and J.L. O'Brien, Science {\bf 325}, 1221 (2009).

\bibitem{cnot1} T.C. Ralph, N.K. Langford, T.B. Bell, and A.G. White, Phys. Rev. A
{\bf 65}, 062324 (2002).

\bibitem{cnot2} H.F. Hofmann and S. Takeuchi, Phys. Rev. A {\bf 66}, 024308
(2002).

\bibitem{cnot3} J.L. O'Brien,  G.J. Pryde,  A.G. White,  T.C. Ralph and D. Branning, Nature {\bf 426}, 264 (2003).

\bibitem{pete} P.J. Shadbolt, M.R. Verde, A. Peruzzo, A. Politi, A. Laing, M. Lobino, J.C.F. Matthews, M.G. Thompson and J.L. O'Brien, arXiv:1108.3309.

\bibitem{matthews} J.C.F. Matthews,  A. Politi,  A. Stefanov, and J.L. O'Brien, Nature Photonics {\bf 3}, 346 (2009).

\bibitem{HOM} C.K. Hong, Z.Y. Ou, and L. Mandel, Phys. Rev. Lett. {\bf 59}, 2044 (1987).

\bibitem{ob_prl} J.L. O'Brien, G.J. Pryde, A. Gilchrist, D.F.V. James, N.K. Langford, T.C. Ralph, and A.G. White, Phys. Rev. Lett. {\bf 93}, 080502 (2004).

\bibitem{maxlike} D.F.V. James, P.G. Kwiat, W.J. Munro, and A.G. White, Phys. Rev. A
{\bf 64}, 052312 (2001). 

\bibitem{joel} J.J. Wallman and S.D. Bartlett, arXiv:1111.1864.

\bibitem{review} S.D. Bartlett, T. Rudolph and R.W. Spekkens, Rev. Mod. Phys. {\bf
79}, 555 (2007).

\bibitem{costa} F. Costa, N. Harrigan, T. Rudolph and C. Brukner, New J. Phys. {\bf
11} 123007 (2009).

\end{thebibliography}
\end{document}